\documentclass[12pt,preprint]{aastex}

\usepackage{natbib}
\begin{document}

\title{A Ring of C$_2$H in the Molecular Disk Orbiting TW~Hya} 

\author{Joel H.\ Kastner\altaffilmark{1}, Chunhua Qi\altaffilmark{2}, Uma Gorti\altaffilmark{3}, Pierre Hily-Blant\altaffilmark{4,5}, Karin Oberg\altaffilmark{2}, Thierry Forveille\altaffilmark{4}, Sean Andrews\altaffilmark{2}, David Wilner\altaffilmark{2}}


\altaffiltext{1}{Chester F. Carlson Center for Imaging Science, School of Physics \& Astronomy, and 
Laboratory for Multiwavelength Astrophysics, Rochester Institute of
Technology, 54 Lomb Memorial Drive, Rochester NY 14623 USA
(jhk@cis.rit.edu)}
\altaffiltext{2}{Harvard-Smithsonian Center for Astrophysics, 60 Garden Street, Cambridge, MA 02138}
\altaffiltext{3}{SETI Institute, 189 Bernardo Ave., Mountain View, CA 94043, USA; NASA Ames Research Center, Moffett Field, CA 94035, USA}
\altaffiltext{4}{Universit\'{e} Grenoble Alpes, Institut de Plan\'{e}tologie et d'Astrophysique de Grenoble (IPAG), F-38000, Grenoble, France; CNRS, IPAG, F-38000, Grenoble, France}
\altaffiltext{5}{Institut Universitaire de France, F-38000, Grenoble, France}

\begin{abstract}
  We have used the Submillimeter Array to image, at $\sim$1.5$''$
  resolution, C$_2$H $N=3\rightarrow2$ emission from the circumstellar
  disk orbiting the nearby ($D = 54$ pc), $\sim$8 Myr-old, $\sim$0.8
  $M_\odot$ classical T Tauri star TW Hya. The SMA imaging
  reveals that the C$_2$H emission exhibits a ring-like
  morphology. Based on a model in which the C$_2$H column density
  follows a truncated radial power-law distribution, we find that the inner edge
  of the ring lies at $\sim$45 AU, and that the ring extends to at
  least $\sim$120 AU.  Comparison with previous (single-dish)
  observations of C$_2$H $N=4\rightarrow3$ emission indicates that the
  C$_2$H molecules are subthermally excited and, hence, that the
  emission arises from the relatively warm ($T\stackrel{>}{\sim}40$ K), tenuous
  ($n<<10^7$ cm${^{-3}}$) upper atmosphere of the disk. Based on these
  results and comparisons of the SMA C$_2$H map with previous submm
  and scattered-light imaging, we propose that the C$_2$H emission
  most likely traces particularly efficient photo-destruction of small
  grains and/or photodesorption and photodissociation of
  hydrocarbons derived from grain ice mantles
  in the surface layers of the outer disk.  The presence of a C$_2$H
  ring in the TW Hya disk hence likely serves as a marker of dust
  grain processing and radial and vertical grain size segregation
  within the disk.
\end{abstract}

\section{Introduction}

Models of protoplanetary disks orbiting low-mass, pre-main sequence (T
Tauri) stars predict that gas-grain processes and high-energy stellar
irradiation are important ingredients in determining disk chemical
composition and dust grain evolution \citep[][and refs.\
therein]{2014A&A...563A..33W,2014prpl.conf..317D}. These evolutionary
processes, in turn, have profound implications for planet-building
scenarios, since disk dust and gas characteristics --- such as characteristic
dust grain sizes, the
presence or absence of grain ice mantles, and gas-phase molecular
abundances --- should regulate the rate of planetesimal buildup and the
compositions of bodies ranging from comets to giant planet envelopes
\citep[e.g.,][and refs.\ therein]{2014prpl.conf..317D}.

Interferometric imaging of molecule-rich disks orbiting nearby T Tauri
star/disk systems --- whose archetype is TW Hya \citep[$D = 54$ pc,
age $\sim$8 Myr;][]{2008hsf2.book..757T,2014A&A...563A.121D} ---
provides key insight into these and other planet formation processes
\citep[e.g.,][]{2004ApJ...616L..11Q,2008ApJ...681.1396Q,2011ApJ...727...85H,2012ApJ...757..129R,2012ApJ...744..162A}. Recently,
for example, imaging with the Atacama Large Millimeter Array (ALMA)
established that N$_2$H$^+$ emission from the TW Hya disk displays a
central hole of radius $\sim$30 AU \citep{2013Sci...341..630Q}. On the
basis of simple gas-phase chemistry, \citet{2013Sci...341..630Q}
interpreted the $\sim$30 AU outer edge of the N$_2$H$^+$ depletion
region as tracing the CO ``snow line,'' i.e., the innermost radius at which CO
freezes out onto dust grains.

In an unbiased molecular line survey of TW Hya spanning the wavelength
range 0.85-1.1 mm, carried out with the Atacama Pathfinder Experiment
(APEX) 12 m telescope, we detected strong C$_2$H $N= 4 \rightarrow 3$
emission \citep{2014ApJ...793...55K}. From analysis of the hyperfine
splitting of this transition, we determined that the emission is
optically thick and either arises from very cold disk layers or is
subthermally excited. We furthermore determined a very large
beam-averaged C$_2$H column density of $\sim$$5\times10^{15}$
cm$^{-2}$, suggesting that, in specific regions of the outer disk, the
abundance of C$_2$H in the gas phase may rival or perhaps exceed that
of CO. As C$_2$H is potentially a product of the photodissociation of
more complex hydrocarbons and dissociative recombinations of molecular
ions with electrons, its radio emission may trace the influence of
stellar irradiation on molecular, protoplanetary disks
\citep{2010ApJ...714.1511H,2010ApJ...722.1607W}. To ascertain the
surface brightness distribution of C$_2$H in the TW Hya disk, which is
viewed nearly pole-on
\citep[$i\approx6^\circ$, assuming a central star mass of $\sim$0.8$M_\odot$;][]{2004ApJ...616L..11Q,2012ApJ...744..162A}, we have carried out
imaging of C$_2$H $N= 3 \rightarrow 2$ emission of TW Hya with the
Submillimeter Array (SMA).

\section{Observations and Data Reduction}

Observations of 262 GHz C$_2$H $N = 3\rightarrow 2$ emission and
adjacent continuum centered at the position of TW Hya were obtained on
2013 April 5 and 2014 April 8 in the SMA's compact (C) and extended
(E) array configurations, respectively, with projected baselines ranging from 6.3 m
to 164.8 m. The correlator setup included the four brightest hyperfine
structure transitions of the C$_2$H $N = 3\rightarrow 2$ rotational transition (which lie in the 262.004--262.067 GHz region; Table 1). The $3_{3,3}-2_{2,2}$  and
$3_{3,2}-2_{2,1}$  transitions were observed in lower sideband correlator 104 MHz
chunk S01, and were blended due to the coarse spectral resolution of
1.86 km s$^{-1}$, while the $3_{4,4}-3_{3,3}$  and $3_{4,3}-3_{3,2}$ transitions were
observed in lower sideband correlator 104 MHz chunk S02 with spectral
resolutions of 0.46 km s$^{-1}$ (C) and 0.23 km s$^{-1}$ (E), which
resolves the two lines (Fig.~\ref{fig:c2h_cont}, bottom panel). The
observing loops used J1037--295 as the gain calibrator, and the
bandpass response was calibrated using observations of 3C279. Flux
calibration was performed using observations of Titan and
Callisto. The derived fluxes of J1037--295 were 0.79 Jy on 2013 April
5 and 0.70 Jy on 2014 April 8. All data were phase- and
amplitude-calibrated with the MIR software package
\footnote{http://www.cfa.harvard.edu/$\sim$cqi/mircook.html}.
Continuum and spectral line maps were then generated and CLEANed using
the MIRIAD software package.

\section{Results and Analysis}

\begin{figure}[htbp]
\begin{center}
\includegraphics[width=6in]{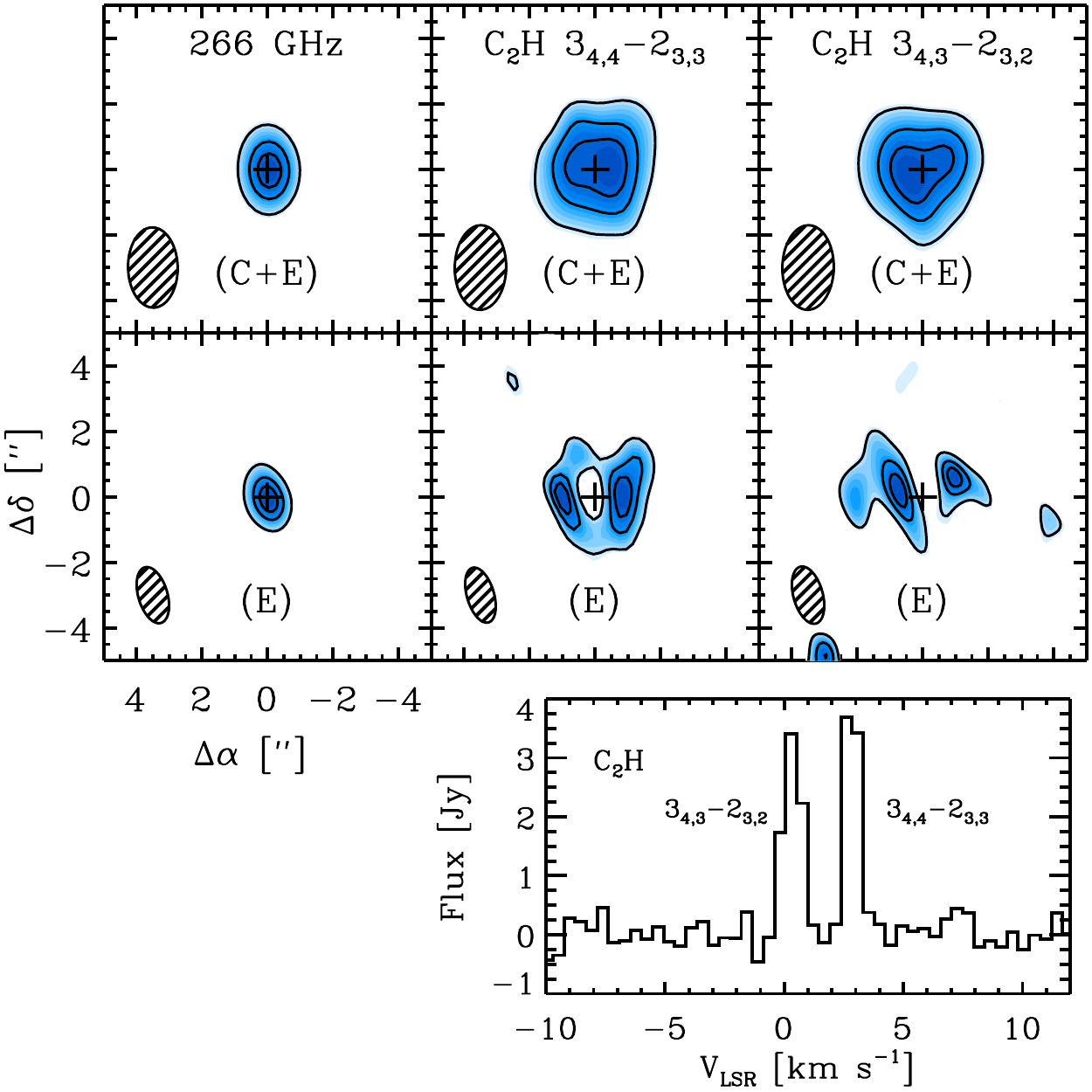}
\caption{{\it Upper two rows of panels}: spectral line maps of the TW Hya disk in 266
  GHz continuum (left) and C$_2$H (3--2) hyperfine lines (center and
  right) synthesized from data obtained in merged compact+extended
  (C+E; top panels) and extended (E; middle panels) SMA antenna
  configurations. In each panel, the shaded oval indicates the FWHM and
  orientation of the synthesized beams, and the cross represents the
  position of TW Hya (J2000 coordinates $\alpha =$ 11:01:51.834, $\delta=$
  $-$34:42:17.149). Contours levels are 50\%, 75\% and 90\% of the
  peak in each image. For the C+E (E)
  configurations, these peak fluxes are: 570 (401) mJy beam$^{-1}$ for the
  266 GHz continuum map, 1.33 (0.49) Jy km
  s$^{-1}$ beam$^{-1}$ for the $3_{4,4}-2_{3,3}$ transition map, and 1.18 (0.41) Jy km
  s$^{-1}$ beam$^{-1}$ for the $3_{4,3}-2_{3,2}$ transition map, with noise levels of 5 (8) and 0.04 (0.08) for
  the continuum and line maps, respectively.
{\it Bottom panel}: spectrum of
  C$_2$H (3--2) hyperfine lines in the 262.005 GHz spectral region.}
\label{fig:c2h_cont}
\end{center}
\end{figure}

In Fig.~\ref{fig:c2h_cont}, we display continuum (266 GHz) and C$_2$H
emission line maps 
for the spectral region encompassing the $3_{4,4}-2_{3,3}$ and
$3_{4,3}-2_{3,2}$ transitions of C$_2$H. From the merged (C+E) data, we measure a 266 GHz continuum
flux of 737$\pm$3 mJy (averaged over both upper and lower sidebands)
and integrated intensities of 3.74 and 3.25 Jy km s$^{-1}$ ($\pm$0.06
Jy km s$^{-1}$), respectively, for the $3_{4,4}-2_{3,3}$ and
$3_{4,3}-2_{3,2}$ transitions (with absolute calibration uncertainties of
$\sim$10\%). These total flux measurements likely
  account for all of the 266 GHz continuum flux and C$_2$H $N = 3\rightarrow2$ 
  line emission from the TW Hya disk, given the angular scales probed by the shortest ($\sim$6 m) SMA baselines at this frequency
  \citep[e.g.,][]{1994ApJ...427..898W}. 
The C$_2$H emission is evidently well
resolved in the merged C+E configuration data (top row of
Fig.~\ref{fig:c2h_cont}). The E configuration map (lower middle
panel of Fig.~\ref{fig:c2h_cont}) reveals that the C$_2$H
($3_{4,4}-2_{3,3}$) emission has a ring-like morphology, with brightened
limbs (ansae) in the E-W direction. 
In the C$_2$H ($3_{4,3}-2_{3,2}$) E configuration 
emission-line map (lower right panel), only the ansae are apparent.

The inner radius of the ring of C$_2$H ($3_{4,4}-2_{3,3}$) emission in
the TW Hya disk, as imaged by the SMA, appears to be larger than that
of the ring of N$_2$H$^+$ (4--3) emission as imaged by ALMA
\citep{2013Sci...341..630Q}. This comparison --- which
  we further explore below (\S 3.2, 4.1) --- suggests that despite their
morphologically similar surface brightness distributions,
C$_2$H and N$_2$H$^+$ may trace different physical and/or chemical
processes in the disk.  The continuum emission is only marginally
resolved in these maps, consistent with previous mm-wave
interferometric imaging observations that established that the
continuum emission from the TW Hya disk is quite compact relative to
the disk molecular line emission \citep[continuum and CO outer radii
of $\sim$60 AU and $\sim$200 AU, respectively;
e.g.,][]{2012ApJ...744..162A}.

\subsection{Excitation of C$_2$H}

Our analysis of the C$_2$H $N= 4 \rightarrow 3$ hyperfine
emission complex had previously indicated that the C$_2$H emission
either arises from very cold (midplane) disk layers, or that the
C$_2$H molecules are subthermally excited
\citep{2014ApJ...793...55K}. The latter interpretation would suggest
that the relative abundance of C$_2$H
peaks high in the disk and, hence, is more consistent with the chemical simulations of protoplanetary disks
presented by \citet{2010ApJ...722.1607W}, in which the relative C$_2$H
abundance peaks at relative heights $z/R \sim 0.3$.

\begin{figure}[htbp]
\begin{center}
\includegraphics[width=4in]{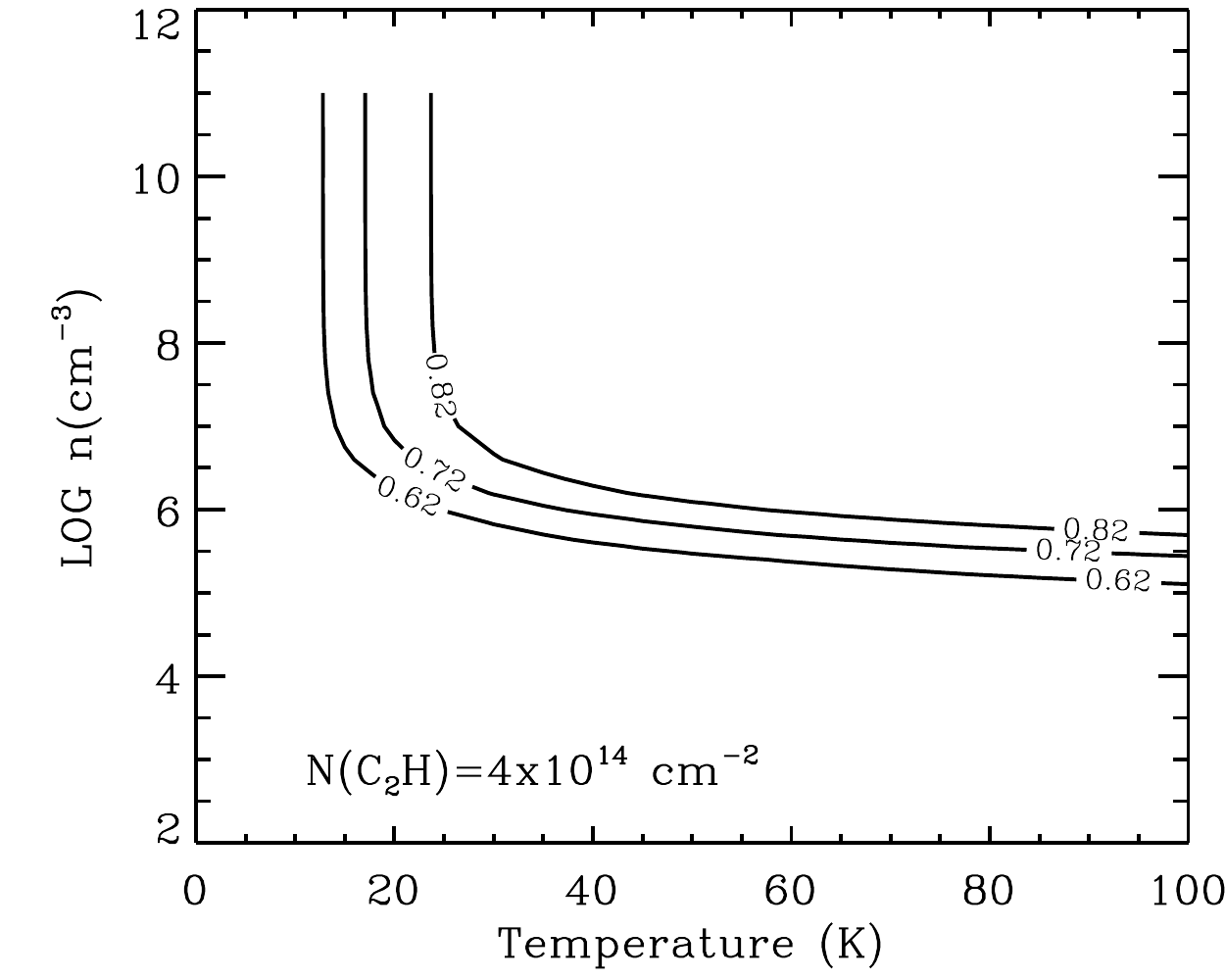}
\caption{RADEX calculations of C$_2$H (4--3/3--2) ratios (see \S 3.1)
  as functions of disk H$_2$ number density ($n$) and gas kinetic temperature
  ($T$) for an assumed C$_2$H column density of $4\times10^{14}$
  cm$^{-2}$ (corresponding to a radial position of $\sim$60 AU, in the surface layer model; \S 3.2) and linewidth 0.7 km s$^{-1}$. The contours indicate the
  values of $n$ and $T$ needed to match the observed C$_2$H
  (4--3/3--2) line ratio (0.72) and its approximate range of
  uncertainty ($\sim \pm 0.1$).}
\label{fig:ratios}
\end{center}
\end{figure}

To investigate whether and how the ratio of C$_2$H $N= 4 \rightarrow
3$ to $N= 3 \rightarrow 2$ line intensities as measured with APEX
\citep[2.7 Jy km s$^{-1}$ for the 349.3375 GHz transition;][]{2014ApJ...793...55K} and
SMA (3.74 Jy km s$^{-1}$ for the 262.0042 GHz transition; \S 3), respectively, constrains
the physical conditions (hence height) of the layers
of the disk from which the C$_2$H emission originates, we constructed
a grid of models with the web-based RADEX
tool\footnote{http://www.sron.rug.nl/\~vdtak/radex/index.shtml }
\citep[][]{2007A&A...468..627V}.  RADEX calculates the emergent line flux under the local velocity gradient approximation, employing the same molecular
  data as used for the morphological analysis in \S 3.2. The resulting model C$_2$H (4--3)/(3--2) line
  ratios were then compared with the observed ratio, i.e., (4--3)/(3--2)
  $\approx$ 0.72. The results, which are
illustrated in Fig.~\ref{fig:ratios}, indicate that the observed
(4--3)/(3--2) line ratio can be reproduced in two specific regimes of
physical conditions: (1) $T \sim 15$--25 K and $\log{n} > 7$, or (2)
$T > 40$ K and $\log{n} \sim 5$--6. The latter conditions correspond
to subthermal excitation of C$_2$H; specifically, the critical
densities for excitation of the 3--2 and 4--3 transitions of C$_2$H
for a gas kinetic temperature $T_k = 30$ K are $n_{cr}\sim$0.6$\times10^7$ cm$^{-3}$ and
$\sim$2$\times10^7$ cm$^{-3}$ respectively (and the values of $n_{cr}$
increase for smaller assumed values of $T_k$).

In our tests of models that can reproduce the ring-like morphology of
the C$_2$H emission (\S 3.2), we find that models in which the C$_2$H
is assumed to lie at relatively deep, dense layers of the disk
consistently overpredict the observed (4--3)/(3--2) line ratio by a
factor $\sim$2 or more, whereas models in which the C$_2$H lies
high in the disk surface layers can reproduce the observed
(4--3)/(3--2) line ratio to within $\sim$30\%.
Furthermore, the very low excitation temperature inferred from the C$_2$H $N= 4
\rightarrow 3$ modeling \citep[$T_{ex}\sim6$ K;][]{2014ApJ...793...55K} is 
inconsistent with the observed (4--3)/(3--2) line ratio, under
conditions of thermal excitation (Fig.~\ref{fig:ratios}). Hence, it appears
that the C$_2$H is confined to low-$n$ (surface) disk
layers, where it is subthermally excited.

\subsection{Modeling the C$_2$H emission morphology}

To constrain the dimensions of the C$_2$H ring in the TW Hya disk, we
modeled the SMA C$_2$H data via the same methodology used to analyze
the ALMA N$_2$H$^+$ (4--3) data \citep[][]{2013Sci...341..630Q}. This
methodology utilizes a model for the TW Hya disk \citep[described in
detail
in][]{2004ApJ...616L..11Q,2006ApJ...636L.157Q,2013Sci...341..630Q} in
which the non-local thermodynamic equilibrium, accelerated 2D Monte
Carlo radiative transfer code RATRAN \citep{2000A&A...362..697H}
serves to calculate the molecular excitation.  The data for the
excitation calculations were drawn from \citet{2005A&A...432..369S},
supplemented with H$_2$ collisional rate coefficients from
\citet{2012MNRAS.421.1891S} and electron impact rates appropriate for C$_2$H from
A. Faure (2014, private comm.). As in the \citet{2013Sci...341..630Q}
treatment of N$_2$H$^+$, the C$_2$H emission is assumed to be produced
within a vertical disk layer, of constant molecular abundance, whose
boundaries are expressed in terms of vertically integrated hydrogen
column densities as measured from the disk surface. The emitting
molecular layer is confined to a ring with inner and outer edges
$R_{in}$ and $R_{out}$, respectively, with a column density that
follows a power-law distribution, i.e., $N \propto R^{p}$.

As noted earlier (\S 3.1), comparison of the SMA and APEX measurements
indicates that the C$_2$H molecules are subthermally excited and,
hence, reside in tenuous disk surface layers.
In applying the foregoing ring-like molecular density distribution to
model the TW Hya C$_2$H emission, we therefore focused on simulations in which the
C$_2$H is confined to a thin surface layer. The specific models described here,
which are representative of this family of simulations,
have upper and lower vertical C$_2$H boundaries at H$_2$ column densities of
$5\times10^{18}$ cm$^{-2}$ and $8\times10^{18}$ cm$^{-2}$,
respectively. 

\begin{figure}[htbp]
\begin{center}
\includegraphics[width=6in]{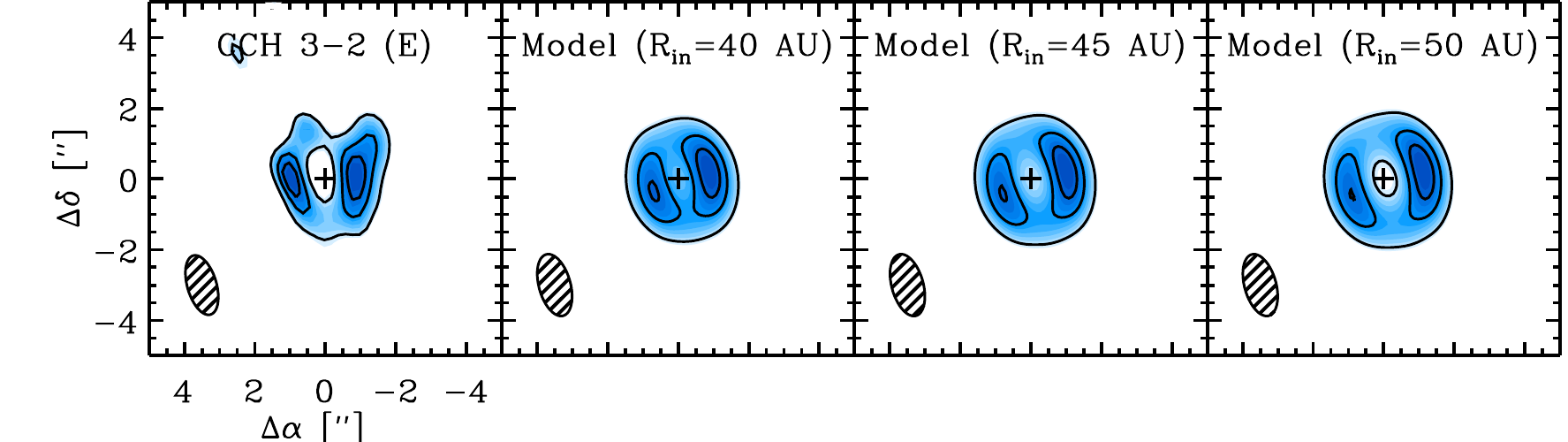}
\caption{Comparison of observed (left panel) vs.\ model (remaining
  panels) maps of C$_2$H emission from the TW Hya disk. In each panel,
  the shaded oval indicates the FWHM and orientation of the
  synthesized or model beam, and the cross represents the position of
  the central star. The inner radii of the model C$_2$H rings
  are indicated in each panel. All models displayed here have the same
  (best-fit) values of outer radius (120 AU) and density power-law
  index ($-1.8$).}
\label{fig:c2h21e_3holes}
\end{center}
\end{figure}

With this vertical distribution fixed, we performed fits to the combined (C+E) visibilities
for the resolved C$_2$H hyperfine lines, with the ring inner and outer
ring edges $R_{in}$, $R_{out}$ and the density power-law index $p$
left as free parameters. We find that the best-fit model,
obtained by minimizing $\chi^2$ for the weighted residuals in the visibilities, has
$R_{in}=45$ AU, $R_{out}=120$ AU, and $p = -1.8$. Based on a
comparison of models with the foregoing values of $R_{out}$ and
$p$ fixed but $R_{in}$ allowed to vary, we find that only models with
$R_{in}$ in the range 40--50 AU well reproduce the observations
(Fig.~\ref{fig:c2h21e_3holes}). The outer radius $R_{out}$ is considerably less well
constrained, due to degeneracy with the value of $p$ during the model
fitting \citep[this degeneracy between $R_{out}$ and $p$ is a
well-known aspect of power-law disk models;
e.g.,][]{1996ApJ...464L.169M}. We note that these results for the
radial distribution of C$_2$H within the disk (i.e., the best-fit
values for $R_{out}$, $R_{out}$, and $p$) are not sensitive to the
assumed vertical placement of the layer from which the C$_2$H emission
arises.

\begin{figure}[htbp]
\begin{center}
\includegraphics[width=3in]{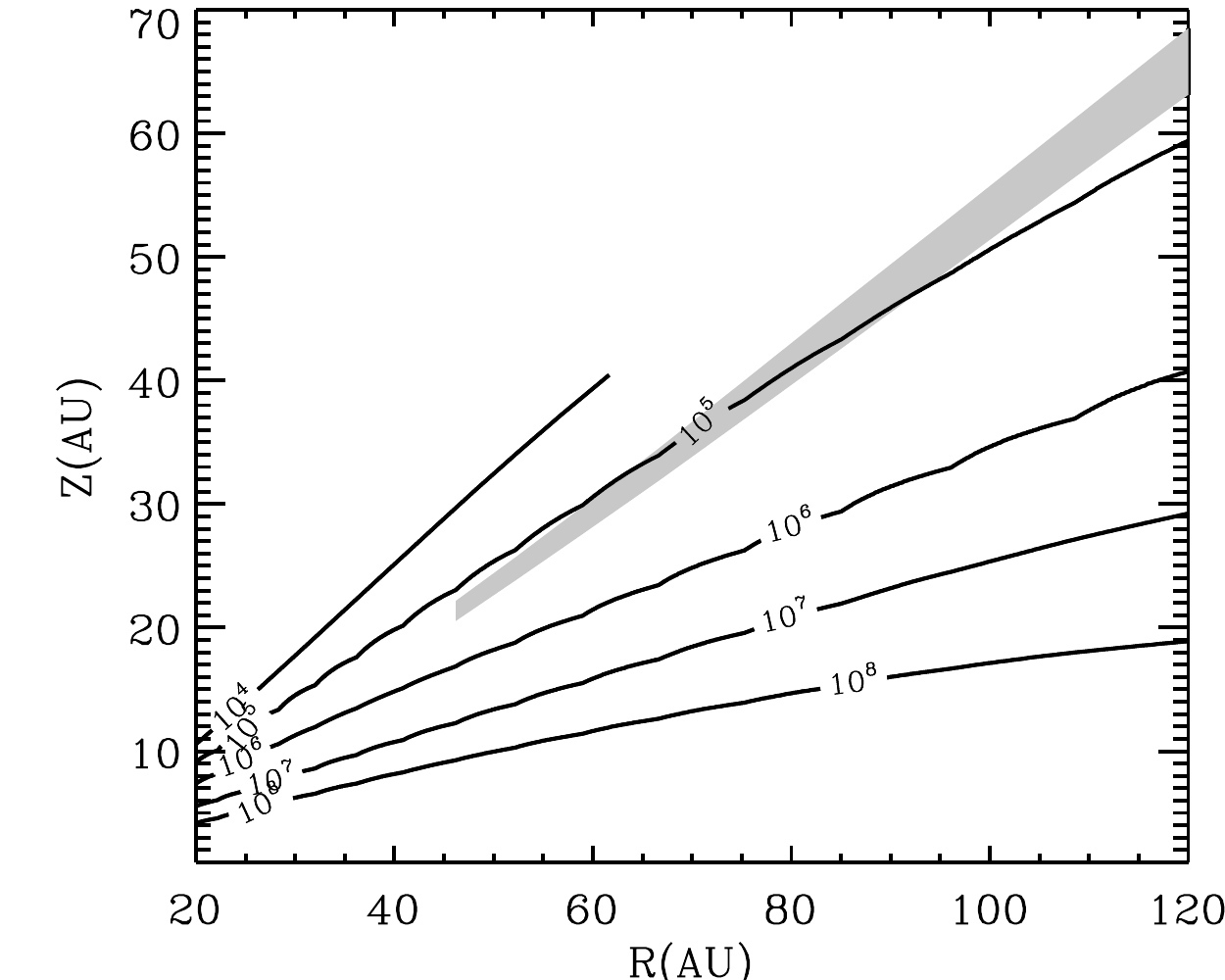}
\includegraphics[width=3in]{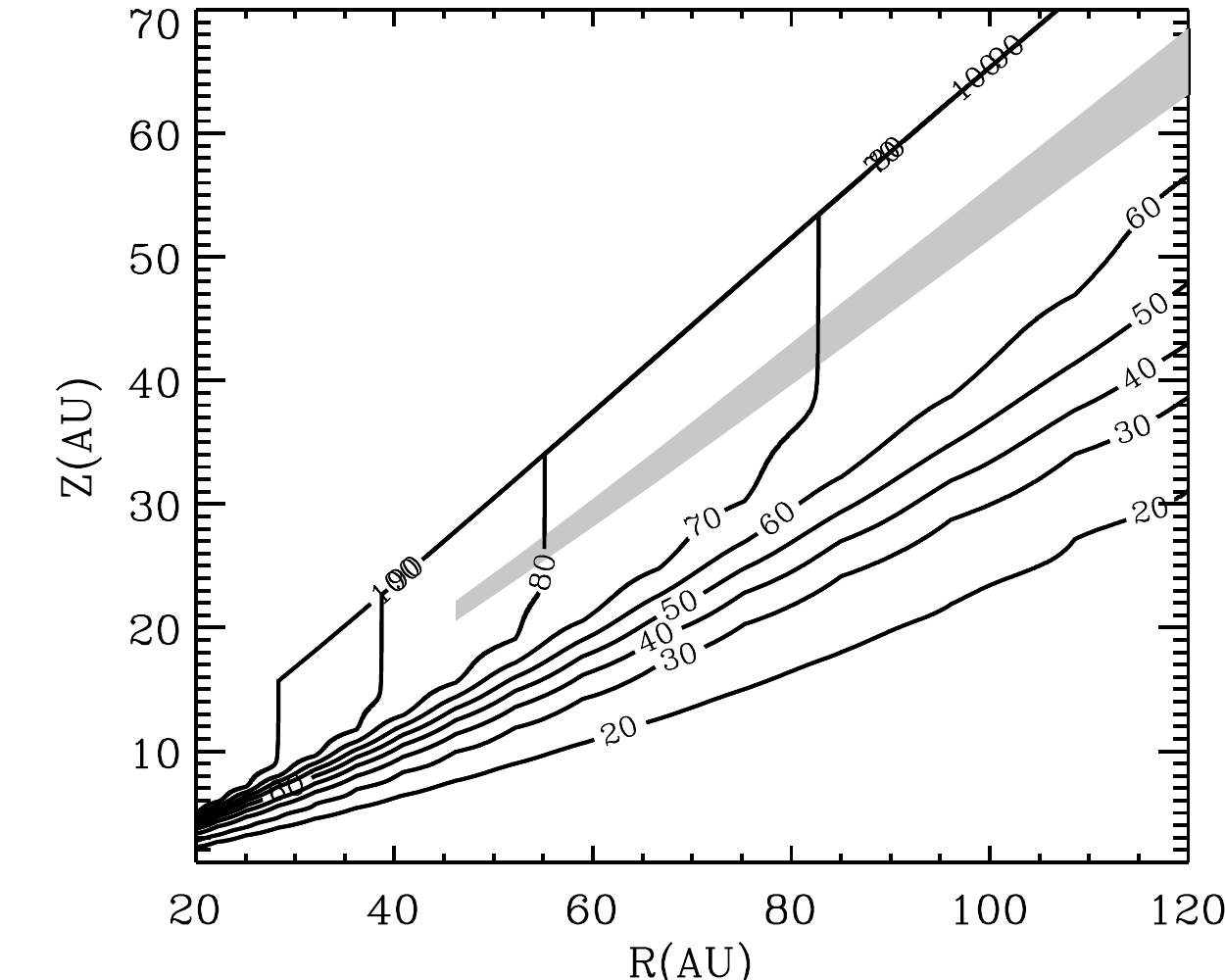}
\caption{Distribution of C$_2$H for the best-fit model, overlaid as
  grey shading on contour plots of disk density ({\it left}, in cm$^{-3}$) and
  gas kinetic temperature ({\it right}, in K).}
\label{fig:modelplot}
\end{center}
\end{figure}

The distribution of C$_2$H in the best-fit surface layer model is
shown schematically in Fig.~\ref{fig:modelplot}. This distribution is
qualitatively similar to that obtained in the simulations by
\citet{2010ApJ...722.1607W}, but lies at lower $n$ (in our model, the
C$_2$H emitting layer corresponds to a disk density $n\approx10^5$
cm$^{-3}$). The column density and approximate relative abundance of
C$_2$H (with respect to H$_2$) are $N_{C_2H}=8.4\times10^{14}$
cm$^{-2}$ and $X_{C_2H}\sim10^{-5}$, respectively, near the inner
emission boundary at $R \approx 45$ AU and $z \approx 20$ AU. The
best-fit value of $N_{C_2H}$ is similar to, though somewhat smaller
than, that estimated from measurements of C$_2$H 4--3 with APEX
\citep{2014ApJ...793...55K}. The value $N_{C_2H}=8.4\times10^{14}$
cm$^{-2}$ likely represents a lower limit on the C$_2$H column
density at $R \approx 45$ AU, given that the C$_2$H emitting layer could lie somewhat
deeper in the disk without violating the constraints imposed by
considerations of excitation (\S 3.1). The inferred value of
$X_{C_2H}$ suggests that the relative abundance of C$_2$H may rival
that of CO in the gas phase, in this region of the disk \citep[see,
e.g.,][]{2013Sci...341..630Q}. The result for $X_{C_2H}$ is highly
uncertain, however, given that the model values of $n$ and $T$ are
poorly constrained in the tenuous disk layers at high $z/R$, and that
the precise C$_2$H emitting layer height and thickness is somewhat
arbitrary.

\section{Discussion}

\subsection{Radial distribution of C$_2$H}

The modeling described in \S 3.1 (and illustrated in
Fig~\ref{fig:modelplot}) indicates that the inner edge of the C$_2$H
ring imaged by the SMA, $R_{in} \approx 45$ AU, lies roughly midway
between the inner edge of N$_2$H$^+$ emission as imaged by ALMA
\citep[$\sim$30 AU;][]{2013Sci...341..630Q} and the outer radius of
submm continuum emission \citep[$\sim$60 AU, as mapped at 870 $\mu$m
and 806 $\mu$m with the SMA and ALMA,
respectively;][]{2012ApJ...744..162A,2013Sci...341..630Q}.  The
N$_2$H$^+$ emission inner cutoff likely traces the radius at which CO
freezes out onto grain surfaces \citep{2013Sci...341..630Q}, whereas
the submm continuum outer cutoff marks the radius at which the large
(radius $\stackrel{>}{\sim}$1 mm) dust grain population drops
precipitously. These two markers (at 30 AU and 60 AU) are well defined
in the respective interferometric data, and neither disk radius is
consistent with the C$_2$H inner hole dimensions
(Fig.~\ref{fig:c2h21e_3holes}). The location of the C$_2$H emission
inner cutoff with respect to the CO snow line --- and the persistence
of C$_2$H emission at disk radii well beyond the drop in large-grain
surface density --- suggests that the C$_2$H abundance is
enhanced in radial zones where small, ice-coated grains are
plentiful. Indeed, at radii $>60$ AU, such small grains likely
dominate the disk's dust grain mass, based on the comparison between
the aforementioned submm imaging and the extended ($\sim$200 AU
radius) scattered-light ``halo'' seen in optical and near-infrared
coronagraphic imaging with the {\it Hubble Space Telescope}
\citep[e.g.,][see \S 4.2.3]{2013ApJ...771...45D}.

\subsection{C$_2$H  production mechanisms}

In the following, we consider three potential processes that might
enhance the abundance of C$_2$H in the surface layers of the TW Hya
disk beyond $\sim$45 AU: {\it (i)} pure gas-phase production of
C$_2$H, involving significant destruction of CO; {\it (ii)} X-ray
photodesorption of organic ices on grains; and {\it (iii)}
photodestruction of very small, C-rich grains and PAHs by UV and
X-rays. As we describe below, mechanisms {\it (ii)} and {\it (iii)}
may be acting in concert to produce the large inferred values of
$N_{C_2H}$.

\subsubsection{Gas-phase production}

A large number of pure gas-phase processes could be
responsible for production of C$_2$H \citep[see, e.g., Fig.\ 7
in][]{2010ApJ...714.1511H}. Many of these channels are initiated via destruction of CO and the consequent generation of ionized C \citep[e.g.,][]{2011A&A...526A.163A,2013ApJ...772....5C}. 
Near the disk surface
--- above the layer where CO is photodissociated by stellar UV and
X-ray photons --- C ions and atoms can react with H$_2$ and CH$_3^+$,
respectively, to form C$_2$H$_2^+$, followed by recombination and
dissociation to C$_2$H. This stellar-photon-driven process should
produce a uniform (as opposed to ring-like) C$_2$H column density
distribution over the disk surface, however, unless the C$_2$H-free
region within $\sim$45 AU is shadowed from high-energy stellar
irradiation by material in the inner disk \citep[e.g., in a puffed-up
inner rim at $\sim$4 AU;][]{2012ApJ...744..162A} .

In regions of the disk shielded from UV and X-rays, CR ionization of He results in the generation of free C. If the C$_2$H ring were generated via such a (CR-initiated)
process, then it should be confined to the disk midplane, where high-energy CRs
dominate the ionization rates \citep{2013ApJ...772....5C}. 
Indeed, our preliminary tests of pure gas-phase chemical models
\citep[utilizing a physical structure for the TW Hya disk described in
detail in][]{2011ApJ...735...90G} do yield a ring-like distribution of
enhanced C$_2$H, with inner radius $\sim$50 AU, deep within the
disk. However, this placement of the emitting C$_2$H (near the disk midplane)
appears to be inconsistent with considerations of molecular excitation
(\S 3.1). Furthermore, we find that these pure gas-phase models
underpredict the C$_2$H column densities by at least an order of
magnitude \citep[similar modeling results were obtained
by][]{2010ApJ...714.1511H}. It therefore appears that pure gas-phase
processes fail to explain the presence of the observed bright C$_2$H
ring in the TW Hya disk.

\subsubsection{X-ray photodesorption of organic ices}

In disk environments such as TW Hya's, where primitive organics (e.g.,
CN, HCN and C$_2$H$_2$) are relatively abundant
\citep[][]{2014ApJ...793...55K}, it is likely that more complex
organic molecules form and freeze out onto dust grains. Desorption of
such ice-coated grains by stellar high-energy radiation ---
especially X-rays --- can then be a rich source of dissociation
products of the parent organics \citep{2013MNRAS.433.3440M}. It is not
clear, however, whether an X-ray-dominated mechanism could produce
C$_2$H high in the atmosphere of the outer disk, given that the stellar
X-rays should penetrate relatively deeply within the disk
\citep{2013ApJ...772....5C}. It is also possible that increasing
gas/dust ratios in these more remote disk regions can facilitate the
dredging of small, ice-coated grains to the disk surface, where they
are efficiently irradiated by ice-destroying UV. However, this
UV-driven photodesorption process should result in efficient C$_2$H
production at radii $<$45 AU, i.e., at least to the CO ``snow line''
at $\sim$30 AU \citep[see also][]{2010ApJ...722.1607W}. Hence --- as in the case
of pure gas-phase processes that are driven by stellar irradiation (\S
4.2.1) --- ``shadowing''
of the disk outer regions (by inner disk material) might be
required to explain the fact that the C$_2$H generated by ice
photodesorption is evidently restricted to somewhat larger disk
radii ($\stackrel{>}{\sim}$45 AU).

\subsubsection{Photodestruction of small grains and PAHs}

Dust grain fragmentation/coagulation equilibrium models
\citep[e.g.,][]{2011A&A...525A..11B,2015arXiv150207369G}
predict both a faster rate of grain
growth in the inner disk, where gas densities are high, and more
efficient collisional shattering of grains at large disk radii, where
the gas surface densities are lower.  The combination of the
relatively sharp cutoff of submm emission in the TW Hya disk at
$\sim$60 AU \citep{2012ApJ...744..162A} and the much larger extent of
the disk in scattered near-infrared light \citep{2013ApJ...771...45D}
is consistent with such modeling. This contrast likely reflects a
transition within the disk from efficient dust grain coagulation and
settling to more efficient grain fragmentation and levitation.  

The outer disk environment hence may harbor an enhanced abundance of
very small grains and polycyclic aromatic hydrocarbons (PAHs),
particularly in the disk's upper layers \citep[see, e.g., Fig.\ 7 in][]{2015arXiv150207369G}. 
Once exposed to stellar UV
and X-rays, these small grains and PAHs would undergo photodegradation
to form loose C$_2$H$_2$ groups \citep{1996A&A...305..602A}, which
would then photodissociate to produce the observed C$_2$H. 
  Preliminary calculations, based on a simple scheme whereby UV
  irradiation breaks PAHs into C$_2$H$_2$ units, suggest that to
explain the C$_2$H column density inferred here (i.e., $N_{C_2H}
\sim10^{15}$ cm$^{-2}$), the relative abundance of PAHs per H nucleus in the upper
layers of the disk would need to be perhaps an order of magnitude
larger than the canonical relative PAH abundance in the interstellar medium
\citep[$\sim$$8\times10^{-7}$;][]{2008ARA&A..46..289T}. As noted,
  such an enhanced PAH abundance could be a byproduct of the steeply
  falling gas surface density in the outer disk, which should yield
  disk conditions favorable for the production of small grains. Furthermore, inclusion of
  X-rays in the PAH photodestruction model would lower the required
  PAH abundance.   Hence, in light of
the modeling results presented in (\S 3.1, 3.2), we consider
photodestruction of small grains and PAHs in disk surface layers to be
a plausible source of the observed C$_2$H. The C$_2$H production rates
from irradiated small grains would be further enhanced if organic ices
coat these grains (\S 4.2.2). 

\subsection{Origin of the C$_2$H ring}

We propose the following schematic model to explain the C$_2$H spatial
distribution, and its apparently subthermal excitation, within the TW
Hya disk. Large surface densities are required to form submm-sized
grains, which also drift radially as the disk evolves; such grain
growth and migration manifests itself as a sharp edge in the submm
dust grain distribution \citep{2012ApJ...744..162A}. A corresponding increase
in small grain surface density in the disk upper layers moving
radially outward, evidently beginning at radii as small as $\sim$45 AU, then
provides ``fertile ground'' for UV and soft X-rays generated at TW Hya
\citep[via chromospheric activity and accretion shocks;
e.g.,][]{2002ApJ...567..434K} to efficiently destroy small grains,
PAHs, and/or grain mantles composed of organic ices. These
photodestruction and photodesorption processes generate hydrocarbons
(predominantly C$_2$H$_2$) and, via subsequent photodissociations,
C$_2$H.
Although pure gas-phase reactions (catalyzed by high-energy
stellar irraditation) could also be responsible for C$_2$H production,
it is unclear whether such processes could produce the observed
ring-like C$_2$H morphology at the large abundances and low densities
inferred from the data.

Additional observations and modeling of C$_2$H emission from
  other, nearby protoplanetary disks are necessary to test and expand
  on the mechanisms proposed here to explain the spatial distribution and
  production of C$_2$H in the TW Hya disk. Indeed, few radio
  measurements of C$_2$H toward disks, either interferometric or
  single-dish, have been published thus far, and the scant interferometric
  data that do exist \citep[e.g.,][]{2010ApJ...714.1511H} are not of
  sufficient quality to determine disk C$_2$H emission
  morphologies. The results of our recent single-dish line surveys of
  V4046 Sgr \citep{2014ApJ...793...55K} and LkCa 15
  \citep{Punzi+2015} 
indicate that these two disks are particularly deserving of followup interferometric
  imaging in C$_2$H. In each case, the inferred C$_2$H abundance and
  excitation is similar to that of TW Hya, indicating that both disks
  may also display ring-like C$_2$H surface brightness distributions.

We further note that in PAH-rich, UV-irradiated
  environments such as the Orion Bar, C$_2$H is merely the most abundant of a
  series of gas-phase hydrocarbons that are potentially detectable in the submm, including C$_4$H, c-C$_3$H$_2$,  and c-C$_3$H, among others \citep[e.g.,][]{2015A&A...575A..82C}. Of
  these, only c-C$_3$H$_2$ has thus far  been detected in a circumstellar molecular
  disk \citep[HD 163296;][]{2013ApJ...765L..14Q}. Sensitive searches for
  this molecule and other simple hydrocarbons are warranted for TW Hya and the other two disks just mentioned, to further test the various C$_2$ production scenarios proposed in \S 4.2.

\section{Conclusions}

The SMA imaging reported here reveals that C$_2$H(3--2) emission from
the TW Hya disk exhibits a ring-like morphology with inner and outer
radii of $\sim$1$''$ and $\sim$3$''$, respectively. Radiative transfer
modeling indicates that the inner edge of the C$_2$H ring lies at
$\sim$45 AU and that the ring extends to at least $\sim$120 AU; we
estimate a lower limit on the C$_2$H column density of
$N_{C_2H}\sim10^{15}$ cm$^{-2}$ near the inner edge of the ring. The inferred inner
cutoff of the C$_2$H emission lies roughly midway between the CO
``snow line'' at $\sim$30 AU (as inferred from N$_2$H$^+$ emission
line imaging) and the sharp decline in submm-sized dust grains at
$\sim$60 AU (as inferred from submm continuum emission).
Comparison with single-dish
observations of C$_2$H(4--3) suggests that the C$_2$H emission
emanates from tenuous upper layers of the disk (characterized by disk
heights $z/R \sim 0.5$ and gas densities $n << 10^7$ cm$^{-3}$), where
the molecules are subthermally excited. While we cannot rule out the
possibility that the C$_2$H arises in colder ($T\sim20$ K),
higher-density regions of the disk, it is more difficult for such a
model to explain the ensemble of observations obtained to date.

We propose that the C$_2$H emission most likely traces particularly
efficient photo-destruction of small grains and/or photodesorption and
photodissociation of hydrocarbons in the surface layers of the outer
disk. The presence of a C$_2$H ring in the TW Hya disk therefore may
be a byproduct of dust grain processing and migration, which is in
turn associated with recent or ongoing giant planet building interior
to the ring.  Given such a scenario, it remains to explain why the
inner cutoff in C$_2$H at $\sim$45 AU evidently lies interior to the
outer cutoff in the large grain surface density.  These results hence
stress the need for higher-resolution (i.e., ALMA) imaging to confirm
and extend the results presented here for TW Hya, and to establish the
radial and vertical distributions of C$_2$H 
within additional examples of nearby, evolved protoplanetary disks orbiting actively
accreting T Tauri stars \citep[e.g., V4046 Sgr, MP Mus, T
Cha, LkCa 15;][]{2014A&A...561A..42S,Punzi+2015}.

\acknowledgments{\it The Submillimeter Array is a joint project between the Smithsonian
Astrophysical Observatory and the Academia Sinica Institute of Astronomy and
Astrophysics and is funded by the Smithsonian Institution and the Academia
Sinica. We gratefully acknowledge Alexander Faure for providing electron impact rates
for C$_2$H, and we thank the anonymous referee for helpful comments and suggestions. This research is supported by National Science
  Foundation grant AST-1108950 to RIT and NASA Origins of Solar
  Systems grant NNX11AK63 to SAO. }


\newpage

\begin{table}[ht!]
\begin{center}
\caption{\sc SMA Correlator Setups for TW Hya C$_2$H Observations}
\begin{tabular}{ccc}
\hline
\hline
C$_2$H transition & Frequency & Resolution \\
  & (GHz) & (km s$^{-1}$) \\
\hline
$N = 3\rightarrow2$, $J=7/2\rightarrow 5/2, F=4\rightarrow 3$ ($3_{4,4}-3_{3,3}$) & 262.004266 & 0.46 (C), 0.23 (E) \\
$N = 3\rightarrow2$, $J=7/2\rightarrow 5/2, F=3\rightarrow2$ ($3_{4,3}-3_{3,2}$) & 262.0064034 & 0.46 (C), 0.23 (E) \\
$N = 3\rightarrow2$, $J=5/2\rightarrow 3/2, F=3\rightarrow 2$ ($3_{3,3}-2_{2,2}$) & 262.0648433 & 1.86 \\
$N = 3\rightarrow2$, $J=5/2\rightarrow 3/2, F=2\rightarrow 1$ ($3_{3,2}-2_{2,1}$) & 262.0673312 & 1.86 \\
\hline
\end{tabular}
\end{center}
\end{table}%

\end{document}